\documentclass[conference]{IEEEtran}
\IEEEoverridecommandlockouts
\usepackage{amsmath,amssymb,amsfonts}
\usepackage{algorithmic}
\usepackage{graphicx}
\usepackage{textcomp}
\def\BibTeX{{\rm B\kern-.05em{\sc i\kern-.025em b}\kern-.08em
    T\kern-.1667em\lower.7ex\hbox{E}\kern-.125emX}}
\usepackage[backend=biber,isbn=false,doi=false,eprint=false,style=ieee,citestyle=numeric-comp]{biblatex}
\usepackage{pgfplots}
\pgfplotsset{compat=1.17}

\begin{document}

\title{%
\parbox{\textwidth}{%
  \normalfont\footnotesize\raggedright
  \textcopyright~2026 IEEE. Personal use of this material
  is permitted. Permission from IEEE must be obtained
  for all other uses, in any current or future media,
  including reprinting/republishing this material for
  advertising or promotional purposes, creating new
  collective works, for resale or redistribution to
  servers or lists, or reuse of any copyrighted
  component of this work in other works.
  
  The final published version is available at
  \url{https://doi.org/10.1109/ICCAD69956.2026.11643097}.
}%
\par\vspace{1em}
Region of Attraction Estimation for Linear Quadratic
Regulator, Linear and Robust Model Predictive Control
on a Two-Wheeled Inverted Pendulum%
\thanks{*This work was supported by the French government
through the France 2030 investment plan managed by the
National Research Agency (ANR), as part of the Initiative
of Excellence Université Côte d'Azur under reference
number ANR-15-IDEX-01.}%
}

\author{\IEEEauthorblockN{1\textsuperscript{st} Lorenzo Fici}
\IEEEauthorblockA{\textit{Université Côte d‘Azur} \\
\textit{CNRS, I3S}\\
Sophia Antipolis, France \\
fici@i3s.unice.fr}
\and
\IEEEauthorblockN{2\textsuperscript{nd} Dalim Wahby}
\IEEEauthorblockA{\textit{Université Côte d‘Azur} \\
\textit{CNRS, I3S}\\
Sophia Antipolis, France \\
wahby@i3s.unice.fr}
\and
\IEEEauthorblockN{3\textsuperscript{rd} Alvaro Detailleur}
\IEEEauthorblockA{\textit{ETH Zürich} \\
\textit{IDSC}\\
Zürich, Switzerland \\
adetailleur@student.ethz.ch}
\and
\IEEEauthorblockN{4\textsuperscript{th} Matthieu Barreau}
\IEEEauthorblockA{\textit{KTH Royal Institute of Technology} \\
\textit{Digital Futures}\\
Stockholm, Sweden \\
barreau@kth.se}
\and
\IEEEauthorblockN{5\textsuperscript{th} Guillaume Ducard*}
\IEEEauthorblockA{\textit{Université Côte d‘Azur} \\
\textit{CNRS, I3S}\\
Sophia Antipolis, France \\
ducard@i3s.unice.fr}
}

\maketitle

\begin{abstract}
Nonlinear underactuated systems such as two-wheeled inverted pendulums (TWIPs) exhibit a limited region of attraction (RoA), which defines the set of initial conditions from which the closed-loop system converges to the equilibrium. The RoA of nonlinear and constrained systems is generally nonconvex and analytically intractable, requiring numerical or approximate estimation methods.
This work investigates the estimation of the RoA for a TWIP stabilized under three model-based control strategies: saturated linear quadratic regulator (LQR), linear model predictive control (MPC), and constraint tightening MPC (CTMPC). We first derive a Lyapunov-based invariant set that provides a certified inner approximation of the RoA. Since this analytical bound is highly conservative, a Monte Carlo-based estimation procedure is then employed to obtain a more representative approximation of the RoA, capturing how the controllers behave beyond the analytically guaranteed region.
The proposed methodology combines analytical guarantees with data-driven estimation, providing both a formally certified inner bound and an empirical characterization of the RoA, offering a practical way to evaluate controller performance without relying solely on conservative analytical bounds or purely empirical simulation.
\end{abstract}

\begin{IEEEkeywords}
Region of attraction (RoA), two-wheeled inverted pendulum (TWIP), Lyapunov, Monte Carlo, linear quadratic regulator (LQR), model predictive control (MPC).
\end{IEEEkeywords}

\section{Introduction}
To assess closed-loop stability in nonlinear systems, we use the concept of the region of attraction (RoA), defined as the set of all initial states from which the closed-loop system converges to a desired equilibrium. By describing the portion of the state space where stability is guaranteed, the RoA provides a direct measure of how far the controller can recover from deviations around the equilibrium point. However, estimating the RoA of nonlinear systems is challenging, as the exact region rarely admits an analytical expression and determining the RoA can be “difficult or even impossible”~\cite{khalil_nonlinear_2002}, since Lyapunov-based estimates are generally conservative. As a result, practical methods for RoA estimation, typically rely on conservative analytical bounds or empirical numerical approximations ~\cite{gross_analytic_2022, tellez_estimate_2013, horibe_quantitative_2018}.

Underactuated robotic systems, such as two-wheeled inverted pendulums (TWIPs), are particularly sensitive to these issues. Their nonlinear and inherently unstable dynamics require feedback controllers that can stabilize the robot in the upright position while operating within the limits of the actuators. In such systems, the RoA serves as a metric for evaluating the closed-loop dynamic's stability: a larger RoA implies a wider range of feasible initial conditions from which balance can be recovered. Thus the RoA is critical for the performance evaluation of controllers.

This work focuses on the estimation of the RoA for a TWIP stabilized using three model-based control strategies: LQR, linear MPC, and constraint-tightening MPC (CTMPC). The analysis begins with a mathematically certified inner approximation of the RoA derived from the nonlinear closed-loop dynamics under the LQR control law, which defines an invariant set. We then verify that, within this set, both MPC and CTMPC generate the same control input as the LQR, therefore the same stability guarantees obtained for the LQR also apply to MPC and CTMPC inside the region. 
To mitigate the conservativeness of this analytical bound, a Monte Carlo–based numerical estimation is employed to obtain a more representative approximation of the RoA by simulating a large number of randomly sampled initial conditions, using the analytical inner approximation as the stopping condition. 

% The proposed methodology combines analytical guarantees with simulation-based estimation for the TWIP, providing a structured framework for RoA analysis. This methodology is general but model-dependent, and applying it to a different system requires recomputing the analytical inner approximation based on that system’s dynamics and parameters.

\subsection{Literature Review}
Given the challenges of estimating the RoA in nonlinear systems, a broad range of methods have been proposed in the literature, differing mainly in the methodology, in the system's representation, and in the nature of the guarantees they provide. Early studies addressed explicitly modeled nonlinear systems, where Lyapunov theory provides a rigorous foundation for local stability analysis. 
Gross et al.~\cite{gross_analytic_2022} studied the RoA of a torque-limited simple pendulum controlled by a saturated LQR. Their analytic method, based on the linearized closed-loop dynamics, yields a conservative inner approximation. They also employed a Monte Carlo sampling approach in which convergence is determined by a numerical tolerance on the state norm, providing no formal stability guarantee.
Aguilar Ibáñez et al.~\cite{ibanez_lyapunov-based_2005} studied an inverted pendulum on a cart and derived an inner approximation of the RoA using Lyapunov’s direct method combined with LaSalle’s invariance principle. The proposed Lyapunov-based controller ensures asymptotic convergence to the upright equilibrium, although the convergence time remains very slow. 
Téllez~\cite{tellez_estimate_2013} experimentally assessed the boundary of the RoA for a Kapitza pendulum on hardware. Although this approach can, in principle, recover the true boundaries of the RoA, it is extremely time-consuming, as each individual experiment yields only a single point of the RoA. 
Horibe et al.~\cite{horibe_quantitative_2018} propose a quantitative stability measure for nonlinear unstable systems based on the minimal radius of the RoA. Using inverted pendulum examples, they define this radius as the shortest distance from the equilibrium to the RoA boundary and compare two RoA estimation strategies: a Lyapunov-based method and Monte Carlo sampling. In contrast, we employ a different metric based on the percentage of stable points over simulated initial conditions, and combine a mathematically-certified inner approximation with Monte Carlo sampling to obtain a more representative and formally grounded estimate of the RoA.

While Lyapunov-based and experimental approaches provide valuable insight into system stability, more recent research has focused on reformulating the RoA estimation problem within a convex optimization framework.
Korda et al.~\cite{korda_inner_2013} formulated the RoA estimation problem as a linear program (LP) over measures and solving its finite-dimensional relaxations using semidefinite programming based on linear matrix inequalities (LMIs). Their method produces a hierarchy of polynomial sublevel sets that provide certified inner approximations of the true RoA, with theoretical guarantees that these sets converge to the actual region as the relaxation order increases. However, this approach is applicable only to systems that can be expressed as polynomial dynamics and suffers from an exponential growth in problem size with the state dimension, making it impractical for the TWIP model and for repeated evaluation of different constrained controllers.
Building on this convex optimization framework, Khattabi et al.~\cite{khattabi_convex_2025} extended the LP-based approach to cases where no explicit model of the dynamics is available. Instead of requiring a model, they rely on measured data together with an upper bound on the system’s Lipschitz constant, which defines a state-dependent uncertainty set consistent with all admissible dynamics. Their method reformulates the RoA estimation problem as an LP defined over an implicit set inclusion derived from a finite dataset and a Lipschitz bound on the system’s behavior. This enables the computation of inner and outer RoA approximations directly from data, but the approach still requires partial knowledge about the system, specifically that the dynamics are Lipschitz and that a Lipschitz bound is known or can be estimated. This method offers numerical rather than formal guarantees, as the accuracy of the estimated regions depends on data quality and coverage. Unlike their approach, which relies on measured data together with a Lipschitz bound because the dynamics are assumed unknown, our method exploits the known nonlinear model of the TWIP to generate arbitrarily dense simulations of initial conditions, enabling RoA estimation without the data-availability limitations.

Beyond convex model-based methods, learning-based approaches have emerged to estimate the RoA directly from experimental data. Berkenkamp et al.~\cite{berkenkamp_safe_2016} proposed a probabilistic framework that learns the RoA of nonlinear systems with uncertain dynamics using Gaussian process models. Their method incrementally expands a conservative estimate of the RoA through safe exploration, ensuring that all sampled trajectories remain within the true stable region with high probability. Although this approach does not yield formal guarantees, it demonstrates how learning can be combined with a Lyapunov-based analysis to achieve data-efficient and safety-critical stability estimation.
Complementary to these methods, Sel et al.~\cite{sel_estimation_2025} addressed the conservativeness of analytical RoA estimation by introducing a multi-transformation approach within the Takagi–Sugeno (TS) modeling framework. By applying several coordinate transformations and computing piecewise quadratic Lyapunov functions for each transformed system, their method captures different local stability properties and combines the results into an enlarged overall RoA. This strategy substantially reduces conservativeness compared to traditional single-transformation TS approaches while preserving the computational tractability of LMI-based analysis.

In summary, existing RoA estimation methods typically fall into two categories: those that provide analytical guarantees, such as Lyapunov-based or convex-optimization approaches, and those that provide empirical coverage through Monte Carlo or data-driven sampling. Lyapunov and convex optimization methods rely on restrictive assumptions, such as requiring polynomial or low-dimensional dynamics and the explicit construction of a Lyapunov function, and in practice, they often characterize only a small inner subset of the RoA rather than the full region. Conversely, purely data-driven or experimental methods lack a mathematically certified core; without an analytical guarantee, they cannot ensure that the states labeled as stable truly converge to the equilibrium. 

\subsection{Contributions}
In this work, we use a nonlinear model of the TWIP to combine a Lyapunov-based inner approximation of the RoA with Monte Carlo sampling, providing both a mathematically certified core of stable initial conditions and an empirical characterization of the surrounding region for comparing the LQR, linear MPC, and CTMPC controllers.
The main contributions of this work are summarized as follows:
\begin{enumerate}
    \item a mathematically-certified invariant set derived from the nonlinear closed-loop dynamics of the TWIP under the LQR control law, providing an inner approximation of the region of attraction that guarantees asymptotic stability within its domain and

    \item a Monte Carlo–based numerical simulation developed to empirically characterize the region of attraction of the three controllers considered in this study: 1) saturated LQR, 2) linear MPC, and 3) CTMPC, by identifying initial conditions that reach the invariant set without violating input constraints.
\end{enumerate}

\section{Nonlinear Model and Controller Synthesis}
\begin{figure}
    \centering
    \includegraphics[width=0.85\linewidth]{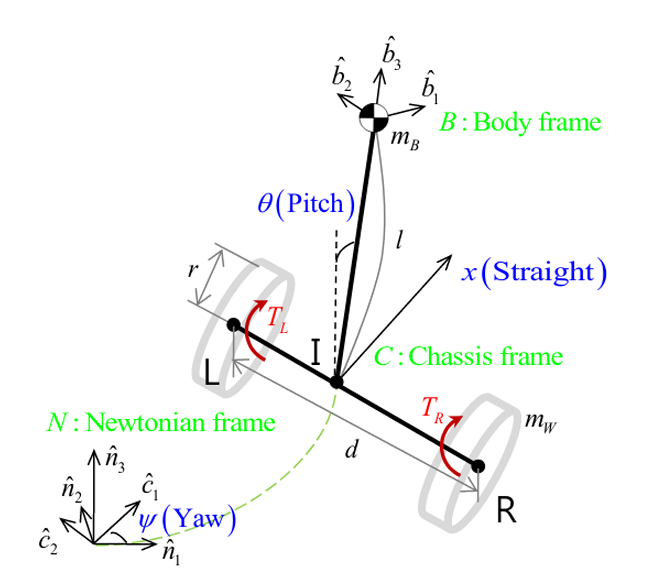}
    \caption{Schematic of the TWIP, indicating physical parameters and coordinate axes. Model parameters are listed in Table~\ref{tab:model_parameters}}
    \label{fig:sigi_schematic}
\end{figure}

To evaluate and compare the performance of different control strategies, we consider the nonlinear dynamic model of the TWIP introduced in~\cite{detailleur_synthesis_2025} and shown in Figure~\ref{fig:sigi_schematic}. 
This model serves as the basis for deriving the model-based controllers analyzed in this work. The equations of motion are expressed as
\begin{equation}
\label{eq:2DOF_dynamics}
\centering
\scalebox{0.9}{$
\begin{aligned}
\ddot{x}_w &= \frac{1}{d_1} \left( aI_O \dot{\theta}^2 \sin\theta - a^2 g \sin\theta \cos\theta + T \left( \frac{I_O}{r} + a\cos\theta \right) \right), \\[4pt]
\ddot{\theta} &= \frac{1}{d_1} \left( -a^2 \dot{\theta}^2 \sin\theta \cos\theta + a\, m_O g \sin\theta - T\left( m_O + \frac{a}{r}\cos\theta \right) \right), \\[4pt]
T &= \frac{K}{R_M} \left( u - K\, i_{gb}\left( \frac{\dot{x}_w}{r} - \dot{\theta} \right) \right)
\end{aligned}
$}
\end{equation}
\noindent where the auxiliary terms are defined as:
\begin{align*}
a &= m_B l, &
I_O &= I_2 + m_B l^2, \\
m_O &= m_B + 2m_W + \frac{J}{r^2}, &
d_1 &= I_O m_O - a^2 \cos^2 \theta,
\end{align*}
and the physical parameters are summarized in Table~\ref{tab:model_parameters}.

\begin{table}
\caption{Model Parameters of the TWIP Robot}
\begin{center}
\begin{tabular}{|c|c|c|c|}
\hline
\textbf{Definition} & \textbf{Symbol} & \textbf{Value} & \textbf{Unit} \\
\hline
Wheel separation & $d$ & 0.10 & m \\
\hline
Wheel axle--COM distance & $l$ & $2.75\times10^{-2}$ & m \\
\hline
Wheel radius & $r$ & 0.04 & m \\
\hline
Body mass & $m_B$ & 0.368 & kg \\
\hline
Mass of each wheel & $m_W$ & 0.02 & kg \\
\hline
Wheel spin inertia & $J$ & $2.25\times10^{-5}$ & kg$\cdot$m$^{2}$ \\
\hline
Gravitational acceleration & $g$ & 9.81 & m/s$^{2}$ \\
\hline
Gearbox ratio & $i_{gb}$ & 49.86 & 1 \\
\hline
Motor constant & $K$ & $1.5\times10^{-3}$ & N$\cdot$m/A \\
\hline
Motor coil resistance & $R_M$ & 12 & $\Omega$ \\
\hline
Body pitch-axis inertia & $I_2$ & $2.17\times10^{-4}$ & kg$\cdot$m$^{2}$ \\
\hline
\end{tabular}
\label{tab:model_parameters}
\end{center}
\end{table}

To design the controllers, the nonlinear dynamics in \eqref{eq:2DOF_dynamics} are linearized around the upright equilibrium point $x=0$, $u=0$, where $x=[x_w,\dot{x}_w,\theta,\dot{\theta}]^\top$, and discretized with sampling time $T_s = 0.01\text{ s}$. All considered controllers are synthesized using the resulting discrete-time linear time-invariant (LTI) model $x_{k+1} = A x_k + B u_k$, with
\begin{equation}
\label{eq:AB_matrices}
\begin{aligned}
A &= 
\left[
\begin{array}{r@{\hspace{12pt}}r@{\hspace{12pt}}r@{\hspace{12pt}}r}
1 & 9.883e{-}3 & -6.524e{-}5 & 4.471e{-}6 \\
0 & 9.768e{-}1 & -1.276e{-}2 & 8.620e{-}4 \\
0 & 6.175e{-}3 & 1.008e{+}0 & 9.780e{-}3 \\
0 & 1.222e{+}0 & 1.573e{+}0 & 9.591e{-}1
\end{array}
\right], \\[6pt]
B &=
\left[
\begin{array}{r}
 6.272e{-}5 \\
 1.240e{-}2 \\
-3.303e{-}3 \\
-6.533e{-}1
\end{array}
\right].
\end{aligned}
\end{equation}

\subsection{Saturated Linear Quadratic Regulator}
The LQR used in this work is adopted as a baseline controller for RoA estimation due to its simplicity and straightforward integration with the nonlinear TWIP model. 
%This approach yields a locally stabilizing control policy that enables the analytical derivation of an invariant set for the nonlinear closed-loop dynamics.
The discrete-time LQR is obtained by solving the infinite-horizon quadratic optimization problem
\begin{equation}
\centering
\scalebox{0.9}{$
\begin{aligned}
\min_{\{u_k\}} \quad & \sum_{k=0}^{\infty} \left( x_k^\top Q x_k + u_k^\top R u_k \right) \\
\text{s.t.} \quad & x_{k+1} = A x_k + B u_k,
\end{aligned}
$}
\label{eq:LQR_cost_discrete}
\end{equation}
where the weighting matrices $Q$ and $R$ 
are defined as:
\begin{equation}
\label{eq:Q_R_matrices}
\centering
\scalebox{0.9}{$
\begin{aligned}
Q &= \mathrm{diag}\!\Big(
10\cdot\!\left(\tfrac{1}{0.1}\right)^{2},\;
5\cdot\!\left(\tfrac{1}{0.2}\right)^{2},\;
50\cdot\!\left(\tfrac{180}{10\pi}\right)^{2},\;
0.1\cdot\!\left(\tfrac{180}{90\pi}\right)^{2}
\Big), \\[3pt]
R &= 250\cdot\!\left(\tfrac{1}{2.5}\right)^{2}.
\end{aligned}
$}
\end{equation}
The resulting saturated state-feedback control law is
\begin{equation}
u_k = \mathrm{sat}\!\left(-K_{\text{LQR}} x_k\right),
\label{eq:LQR_law_discrete}
\end{equation}
where $\mathrm{sat}(\cdot)$ denotes the element-wise saturation operator enforcing the actuator constraint $u_k \in [-2.2,\, 2.2]\,\mathrm{V}$ and the corresponding feedback gain matrix \( K_{\text{LQR}} = [-4.291,\ -8.142,\ -9.271,\ -0.574] \).

Note that due to actuator saturation and the limited validity of the linearization, the resulting controller guarantees only local asymptotic stability rather than global stability.

\subsection{Linear Model Predictive Control}
Linear MPC solves a finite-horizon constrained optimal control problem at each sampling instant based on the linearized model of the system.
\begin{equation}
\centering
\scalebox{0.86}{$
\begin{aligned}
\min_{\substack{\{u_k\}_{k=0}^{N-1}\\ \{\epsilon_k\}_{k=0}^{N}}} \quad &
\sum_{k=0}^{N-1} \left( x_k^\top Q x_k + u_k^\top R u_k + \rho \|\epsilon_k\|_2^2 \right)
+ x_N^\top Q_N x_N + \rho \|\epsilon_N\|_2^2 \\[4pt]
\text{s.t.} \quad &
x_{k+1} = A x_k + B u_k,\qquad k = 0,\ldots, N-1,\\
& A_{\mathcal{X}}\, x_k \le b_{\mathcal{X}} + \epsilon_k,\qquad \epsilon_k \ge 0,\qquad k = 0,\ldots, N-1,\\
& u_k \in \mathcal{U},\qquad k = 0,\ldots, N-1,\\
& A_{\mathcal{X}_f}\, x_N \le b_{\mathcal{X}_f} + \epsilon_N,\qquad \epsilon_N \ge 0.
\end{aligned}
$}
\label{eq:mpc_single}
\end{equation}
The matrices $Q$ and $R$ are chosen equal to those used in the LQR design (\ref{eq:Q_R_matrices}).
The terminal cost \(Q_N\) is chosen as the solution of the discrete-time algebraic Riccati equation (DARE).
The horizon is set to $N = 20$.
The state and input constraints \(\mathcal{X}\) and \(\mathcal{U}\), respectively, are defined as \(\mathcal{X} = \{\, x \in \mathbb{R}^4 \mid |\dot{x}_w| \le 0.5~\mathrm{m/s},\, |\theta| \le 0.75~\mathrm{rad},\, |\dot{\theta}| \le 5.5 ~\mathrm{rad/s}\,\}\) and \(\mathcal{U} = \{\, u \in \mathbb{R} \mid |u| \le 2.2 ~\mathrm{V}\,\}\). The expression \(\Lambda x \le b\) denotes the half-space form.
The terminal constraint set is computed as the maximal positively invariant set under the local LQR feedback.  
Slack variables \(\epsilon_k\) are penalized by $\rho = 10^5$ to allow softening of state and terminal constraints while maintaining feasibility in receding-horizon simulation.
\subsection{Constraint Tightening Model Predictive Control}
Constraint tightening MPC is a robust MPC formulation that ensures constraint satisfaction in the presence of bounded disturbances on the input. Instead of optimizing the real system directly, CTMPC optimizes a nominal trajectory and constrains it to lie inside tightened state and input sets. 
A feedback controller keeps the real system within an invariant tube around the
nominal trajectory, guaranteeing that all possible disturbance realizations
remain feasible.

First, we distinguish the real system, with state variable $x_k$, which is affected by an additive bounded disturbance on the input $w_k\in\mathcal{W}$, where $\mathcal{W} = \{\, B w \mid w \in [-w_{\max}, w_{\max}] \,\}$ with $w_{\max} = 0.075~\mathrm{V}$, from the nominal prediction model, with state variable $z_k$:
\begin{equation}
x_{k+1}=Ax_k+Bu_k+w_k,\qquad
z_{k+1}=Az_k+Bv_k,
\end{equation}
where $v_k$ is the nominal input optimized by the controller.
Then, the error is defined as $e_k:=x_k-z_k$.
A stabilizing state–feedback gain $K_{\text{tube}}$ is used to contract the error dynamics, and the input applied to the system becomes $u_k = v_k + K_{\text{tube}} e_k$. Substituting this into the dynamics yields
\begin{equation}
e_{k+1}=(A+BK_{\text{tube}})e_k+w_k.
\end{equation}
At each sampling instant, the CTMPC controller solves the following finite-horizon optimization problem:
\begin{equation}
\centering
\scalebox{0.88}{$
\begin{aligned}
\min_{\substack{\{v_i\}_{i=0}^{N-1}\\ \{\epsilon_i\}_{i=0}^{N}}} \quad &
\sum_{i=0}^{N-1} \left( z_i^\top Q z_i + v_i^\top R v_i + \rho \, \epsilon_i^2 \right)
\;+\; z_N^\top Q_N z_N + \rho \, \epsilon_N^2 \\[3pt]
\text{s.t.} \quad &
z_{i+1} = A z_i + B v_i, \qquad i = 0,\ldots, N-1,\\
& z_i \in \{ x \mid \Lambda_{\bar{\mathcal{X}}_i} x \le b_{\bar{\mathcal{X}}_i} + \epsilon_i \mathbf{1}_{\bar{\mathcal{X}}_i} \}, \, \epsilon_i \ge 0,\, i = 0,\ldots, N-1
,\\
& v_i \in \bar{\mathcal{U}}_i = \mathcal{U} \ominus K_{\text{tube}} \, \mathcal{F}_i, \qquad i = 0,\ldots, N-1,\\
& z_N \in \{ x \mid \Lambda_{\bar{\mathcal{X}}_f} x \le b_{\bar{\mathcal{X}}_f} + \epsilon_N \mathbf{1}_{\bar{\mathcal{X}}_f}
 \}, \qquad \epsilon_N \ge 0,\\
& z_0 = x.
\end{aligned}
$}
\label{eq:opt_prob_CTMPC}
\end{equation}
The nominal constraints are tightened using the Pontryagin difference \cite{mayne_robust_2006}, denoted by $\ominus$, yielding $\bar{\mathcal{X}}_i = \mathcal{X} \ominus \mathcal{F}_i$ and $\bar{\mathcal{X}}_f = \mathcal{X}_f \ominus \mathcal{F}_N$. The symbols $\mathbf{1}_{\bar{\mathcal{X}}_i}$ and $\mathbf{1}_{\bar{\mathcal{X}}_f}$ denote vectors of ones with the same dimension as $b_{\bar{\mathcal{X}}_i}$ and $b_{\bar{\mathcal{X}}_f}$, respectively.
Here, $\mathcal{F}_i$ denotes the set of all possible state deviation errors at prediction step $i$, and in this CTMPC formulation, in contrast to classical robust tube MPC \cite{mayne_robust_2006}, the tube size changes over time, generating a funnel composed of the sequence of ellipsoids defined as $\mathcal{F}_i = \{\, e_i \mid e_i^\top P_{\text{tube}} e_i \le \delta_i^2 \,\}
$. The tube tightening follows the recursion $\delta_{i+1} = \alpha\,\delta_i + \delta_1$, where $\alpha = 0.815$ is a chosen design parameter that specifies the desired contraction rate of the tube. The term $\delta_1$ represents the bound on the error caused by the disturbance on the first step, $e_1 = w_0$, and since by definition $z_0 = x$, we have $e_0 = 0$, which implies $\delta_0 = 0$. 

The softened constraints in the optimization are written as $z_i \in \bar{\mathcal{X}}_i \oplus \epsilon_i$ and $z_N \in \bar{\mathcal{X}}_f \oplus \epsilon_N$, where $\oplus$ denotes the Minkowski sum, and $\epsilon_i \ge 0$. The slack variables $\epsilon_i$ allow temporary constraint relaxation and are penalized in the cost function to discourage violations through the parameter $\rho = 10^4$.
The weighting matrices \(Q\) and \(R\) are chosen as in the linear MPC formulation~(\ref{eq:Q_R_matrices}) and the horizon is set to $N=20$. \(Q_N\) is the solution to the DARE, defining the terminal cost.
The terminal set \( \mathcal{X}_f \) is computed as the maximal robust positively invariant (RPI) set of the closed-loop dynamics for the disturbance set $\mathcal{W}$.

The matrices $K_{\text{tube}}$, $P_{\text{tube}}$, and the disturbance bound $\delta_1$ are computed offline by solving a semidefinite program (SDP). For the chosen contraction rate $\alpha$, the SDP determines $K_{\text{tube}}$ and a Lyapunov matrix $P_{\text{tube}}$ satisfying $(A + B K_{\text{tube}})^\top P_{\text{tube}} (A + B K_{\text{tube}}) \preceq \alpha\, P_{\text{tube}}$, while $\delta_1$ satisfies $e_1^\top P_{\text{tube}} \, e_1 \le \delta_1^2$.
The resulting values can be found in \cite{fici_comparative_nodate}.
% \begin{align*}
% \delta_1 &= 2.987\times10^{-2} \\[1ex]
% P_{\text{tube}} &=
% \begin{bmatrix}
% 1.912\mathrm{e}{+}8 & 1.210\mathrm{e}{+}7 & 4.049\mathrm{e}{+}6 & 2.325\mathrm{e}{+}5 \\
% 1.210\mathrm{e}{+}7 & 8.779\mathrm{e}{+}5 & 2.636\mathrm{e}{+}5 & 1.691\mathrm{e}{+}4 \\
% 4.049\mathrm{e}{+}6 & 2.636\mathrm{e}{+}5 & 8.644\mathrm{e}{+}4 & 5.071\mathrm{e}{+}3 \\
% 2.325\mathrm{e}{+}5 & 1.691\mathrm{e}{+}4 & 5.071\mathrm{e}{+}3 & 3.258\mathrm{e}{+}2
% \end{bmatrix} \\
% K_{\text{tube}} &=
% \begin{bmatrix}
% 1.545\mathrm{e}{+}4 & 1.438\mathrm{e}{+}3 & 3.821\mathrm{e}{+}2 & 2.848\mathrm{e}{+}1
% \end{bmatrix}
% \end{align*}

\section{Region of Attraction Estimation}
\label{sec:roa_estimation}

This section describes the methodology used to estimate the RoA of the nonlinear TWIP under different control laws. The approach combines an analytical inner approximation based on Lyapunov stability analysis with a data-driven Monte Carlo estimation using the nonlinear model.

\subsection{Analytical Inner Approximation of the RoA}
\label{subsec:invariant_set}
In~\cite{gross_analytic_2022, fici_comparative_nodate}, the Monte Carlo estimation of the RoA relied on a simple stopping condition based on a tolerance on the Euclidean distance between the final state and the origin. However, this criterion does not provide a formal guarantee that the simulated trajectories actually converge to the equilibrium. To address this limitation, we propose an invariant terminal set derived directly from the closed-loop dynamics of the TWIP under the saturated LQR controller. 
Following the approach outlined in the proof of Theorem 4.7 in Khalil’s Nonlinear Systems~\cite{khalil_nonlinear_2002}, we rewrite the dynamics as $x_{k+1} = A_{\text{cl}} \, x_k + g(x_k)$, where $A_{\text{cl}} = A-BK_{\text{LQR}}$ denotes the closed-loop system matrix and $g(x_k)$ collects all nonlinear terms. The linear part is convergent since $A_{\text{cl}}$ is Hurwitz. The nonlinear part is bounded locally as $\|g(x_k)\| < \gamma \|x_k\|$ for all $\|x_k\| < \rho$. By selecting $\gamma$ such that
\begin{align}
\gamma < \frac{-\|P A_{\text{cl}}\| + \sqrt{\|P A_{\text{cl}}\|^2 + \lambda_{\min}(Q)\,\lambda_{\max}(P)}}{\lambda_{\max}(P)},
\end{align}
where P is the solution to the discrete-time Lyapunov equation $A_{\text{cl}}^\top P A_{\text{cl}} - P = -Q,$ with $Q$ equal to the $4\times 4$ identity matrix, then the quadratic Lyapunov function $V(x) = x^\top P x$ decreases for all states satisfying $\|x_k\| < \rho$.

However, the condition $\|x_k\| < \rho$ only guarantees $\dot{V}(x) <0$ and does not define an invariant set. To obtain an invariant set, we select the largest Lyapunov sublevel set fully contained in the ball $\{x_k : \|x_k\| < \rho\}$. From
\begin{align}
\lambda_{\min}(P)\,\|x\|^2 \;\le\; x^\top P x \;\le\; \lambda_{\max}(P)\,\|x\|^2,
\end{align}
every state satisfying $x^\top P x \le \lambda_{\min}(P)\,\rho^2$ also satisfies $\|x\| < \rho$. Thus, the set
\begin{align}
\mathcal{X} = \{\, x \,:\, x^\top P x \le \lambda_{\min}(P)\,\rho^2 \,\}.
\end{align}
is the largest Lyapunov sublevel set contained in the region where $\dot{V}(x) <0$, and is therefore invariant under the LQR closed-loop dynamics. 
% This geometric construction is illustrated in Figure~\ref{fig:methodology}.
% \begin{figure}[t]
%     \centering
%     \includegraphics[width=0.9\linewidth]{Figures/Methodology.png}
%     \caption{Geometric illustration of the analytical inner approximation of the RoA.}
%     \label{fig:methodology}
% \end{figure}

This region represents a mathematically justified invariant set under the LQR control law, obtained from a Lyapunov-based condition that guarantees convergence to the origin. It serves as a stopping condition for the Monte Carlo–based RoA estimation because, once a trajectory enters this set, convergence to the equilibrium is guaranteed.

\subsection{Validity of the Invariant Set for MPC and CTMPC}
\label{subsec:invariant_admissibility}
Since the analytical inner approximation of the RoA is obtained under the assumption that the LQR control law is applied, we first need to ensure that the same condition holds for MPC and CTMPC. As shown in Proposition~5.2.1 of Johansson~\cite{johansson_linear_2024}, 
when the terminal cost $Q_N$ is chosen as the solution of the DARE, the MPC input is identical to the LQR input whenever no constraints are active. Since CTMPC reduces to nominal MPC whenever the tightened constraints are inactive, it also matches the LQR law inside the terminal set.

For both controllers, we then verify that the entire invariant set lies within their respective state, input, and terminal constraints. This is assessed by computing the largest scaling factor \(\alpha^\star\) such that the ellipsoid \(x^\top P x \le \alpha^\star\), where $P$ is the matrix defining the local Lyapunov function, remains fully contained within these constraint sets. For the CTMPC, the same procedure is applied using the tightened constraint sets evaluated at the final prediction step, where the constraints are most restrictive. In both cases, the computed $\alpha^\star$ is larger than the radius of the invariant set, confirming that the invariant set is fully admissible for all controllers.

\subsection{Monte Carlo-based Estimation}
To assess the stabilization capability of the controllers, we estimate their RoA using a Monte Carlo-based numerical procedure. A large number of initial states is randomly sampled from a bounded region of the state space, and for each sample, a nonlinear closed-loop simulation is performed. 

Contrary to approaches that classify convergence based on a numerical tolerance on the state norm \cite{gross_analytic_2022, fici_comparative_nodate}, we use the analytically derived invariant set introduced in Section~\ref{subsec:invariant_set} as a certified stopping condition. Once a trajectory enters this set, convergence to the origin under the LQR control law is mathematically guaranteed. As shown in Section~\ref{subsec:invariant_admissibility}, the same guarantee applies to MPC and CTMPC within this region.

A total of 5000 initial conditions were sampled from the 3D grid defined by $\dot{x}_w \in [-1,1]~\mathrm{m/s}$, $\theta \in [-1.5,1.5]~\mathrm{rad}$, and $\dot{\theta} \in [-1.5,1.5]~\mathrm{rad/s}$, while the cart position was fixed to $x_w = 0$ for visualization purposes. The system was simulated with a sampling time of $0.01\,\mathrm{s}$ over a $20\,\mathrm{s}$ horizon.

\begin{figure}
  \centering
  \includegraphics[width=0.8\linewidth]{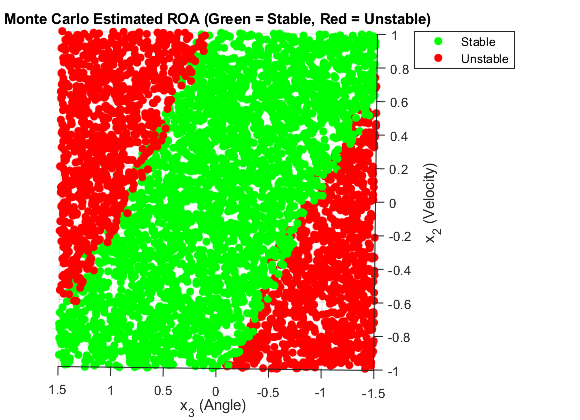}\vspace{0.5em}
  \includegraphics[width=0.8\linewidth]{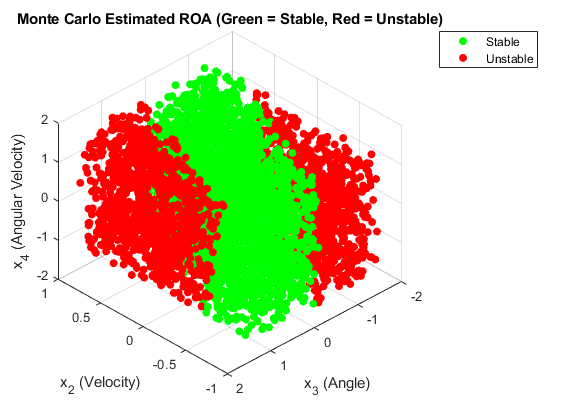}
  \caption{Monte Carlo estimation of the RoA for LQR, MPC, and CTMPC, shown from two perspectives.}
  \label{fig:lqr_roa_mc}
\end{figure}

As shown in Figure~\ref{fig:lqr_roa_mc}, all three controllers yielded the same RoA estimates: $58.48\%$ of the sampled states entered the invariant set, and thus were guaranteed to converge. Since the estimated RoA obtained through Monte Carlo sampling is identical for all three controllers, the most direct conclusion is that their true RoA is likely very similar. This statement is limited by the discrete sampling grid: stability can only be asserted for the specific sampled initial conditions that reached the invariant set, and no claims can be made about the unsampled points between them. Still, within the explored region, the results indicate that LQR, MPC, and CTMPC stabilize essentially the same set of initial conditions, despite the differences in their underlying control laws and the constraints they enforce.

\subsection{Empirical RoA Validation on Hardware}
To provide practical insight into the estimated RoA, hardware experiments were conducted on the TWIP robot using the saturated LQR under reference tracking conditions. The following results aim to qualitatively illustrate the consistency between the estimated RoA and the observed closed-loop behavior.
Two representative experiments were performed using the LQR controller under reference tracking conditions.\footnote{A video demonstration of the experiments is available online:\\
\url{https://www.youtube.com/@GDLab-Research}}

In the first experiment, the system tracks a constant velocity reference of $-0.5\,\text{m/s}$. External input disturbances, applied as short-duration square pulses ($0.02\,\text{s}$) of increasing amplitude (up to $1.4\,\text{V}$) at different time instants, drive the system state toward the boundary of the estimated RoA. The state reaches $[-6.4\,\text{m},\,-0.53\,\text{m/s},\,-0.21\,\text{rad},\,1.5\,\text{rad/s}]^\top$, after which the system successfully recovers and returns to the equilibrium.

In the second experiment, the system tracks a higher velocity reference of $-1.0\,\text{m/s}$. Under the same disturbance conditions, the state is driven beyond the estimated RoA boundary, reaching $[-3.9\,\text{m},\,-1.1\,\text{m/s},\,-0.5\,\text{rad},\,-2.6\,\text{rad/s}]^\top$. In this case, the system fails to recover, leading to loss of stability.

These results are consistent with the estimated RoA and demonstrate the practical applicability of the proposed analysis.

\section{Conclusion}
This work presented a methodology for estimating the RoA and validated its application on a two-wheeled inverted pendulum under three model-based controllers: 1) saturated LQR, 2) linear MPC, and 3) constraint tightening MPC. A Lyapunov-based invariant set was derived directly from the nonlinear closed-loop dynamics, providing a mathematically certified inner approximation of the RoA. As expected, this analytical region is very small, reflecting the conservativeness of Lyapunov-based guarantees in nonlinear constrained systems.

To obtain a more representative picture of the closed-loop behavior, a Monte Carlo–based procedure was used to identify initial states that reach the certified invariant set without violating constraints. This allowed to visualize the initial conditions that successfully converge, providing an empirical approximation of the RoA for each controller. The results show that LQR, linear MPC, and CTMPC achieve the same estimated region of attraction on the TWIP, indicating that the RoAs of the three controllers are probably very similar.

Overall, the proposed framework combines analytical guarantees with empirical characterization, enabling RoA analysis for nonlinear systems when analytical methods are overly conservative and data-driven approaches lack formal guarantees. This methodology is general but model-dependent, and applying it to a different system requires recomputing the analytical inner approximation based on that system’s dynamics and parameters. Furthermore, the inner approximation is inherently local and becomes more conservative in higher-dimensional systems. Moreover, the invariant set is obtained in a single step; computing its predecessors could enlarge the certified region and represent a promising direction for future work.

\section*{Acknowledgment}
%We sincerely thank Dr. Franck Diard, Chief Software Architect at NVIDIA, for financially supporting this work.
This work was financially sponsored by Franck Diard, Chief Software Architect at NVIDIA, through Fondation UniCA.

\addtolength{\textheight}{-12cm}
\printbibliography
%\bibliographystyle{IEEEtran}
%\bibliography{references}
\end{document}